\documentclass{svjour3}                     
\usepackage{graphicx}
\usepackage{pgfplots}
 \pgfplotsset{compat=1.17}

\begin{document}

\title{Towards efficient multilayer network data management
}


\author{Georgios Panayiotou \and Matteo Magnani \and Bruno Pinaud}

\institute{G. Panayiotou \and M. Magnani \at
              InfoLab, Dept. of Information Technology, Uppsala University, Sweden\\         
           \and
           B. Pinaud \at
              Univ. Bordeaux, CNRS, Bordeaux INP, LaBRI, UMR 5800, France\\
}

\date{}

\maketitle

\begin{abstract}
Real-world multilayer networks can be very large and there can be multiple choices regarding what should be modeled as a layer.
Therefore, there is a need for their effective storage and manipulation.
Currently, multilayer network analysis software use different data structures and manipulation operators. We aim to categorize operators in order to assess which structures work best for certain operator classes and data features.
In this work, we propose a preliminary taxonomy of layer and data manipulation operators. 
We also design and execute a benchmark of select software and operators to identify potential for optimization.
 \keywords{Multilayer networks \and Multiplex networks \and Data management}
 \end{abstract}


\section{Introduction}
\label{intro}

Multilayer networks 
have become increasingly popular across various disciplines for representing, manipulating and analyzing complex systems, such as brain~\cite{DeDomenico2017} and ecological~\cite{Timoteo2018} networks. Typical applications also include social networks, which often consist of millions of actors and associated relationships when collected from online sources~\cite{SMLN}. With large multilayer network data availability increasing, an open challenge is to provide a framework for effective storage and access, which will help reducing data preprocessing costs~\cite{simplification} and support development of interactive analysis systems~\cite{VizBook}. 


Using established database management systems for multilayer network data storage is not ideal, as the state-of-the-art standards, i.e. relational and graph database systems, lack methods to represent, manipulate and analyze multilayer network data.
On the other hand, there is a lack of consensus on the appropriate data structure for multilayer network data storage, as quite a few different alternatives exist within analysis and visualization libraries. 
More importantly, there is an imbalance between scalability of a software's underlying data structures and process efficiency; depending on the chosen data model, the performance of even simple layer manipulation operators vastly differs (cf.~Sect.~\ref{bench}). 

Despite increasing interest in multilayer networks, work on multilayer network data management remains scarce. Previous work by the data engineering community on heterogeneous information networks~\cite{HIN} neither includes the concept of layer nor has looked into specialized storage solutions to optimize layer operations. In the recently proposed multilayer graphs, an extension for property graphs~\cite{SIGMOD22}, the term layer is used, but instead refers to levels of nesting in the network. Also recently, a data mining approach was proposed, based on converting EER-diagrams into multilayer networks~\cite{DKE22}; however, the previous does not directly address the impact of different data structures for multilayer networks.

With the previous issues in mind, this work provides basis for a data management framework natively supporting multilayer networks, by considering storage, access and manipulation of both vertices and edges in the network, and the layers themselves.
We propose a taxonomy covering and extending layer and data manipulation operators found in popular multilayer network software~\cite{MuxViz,multinet,Pymnet,Tulip}. 
Finally, we provide a benchmark of select operators' performance on currently available libraries
and database models able to represent a multilayer network.
We aim to both compare alternative data management approaches from a scalability and efficiency perspective, and spotlight processes in need of further research. 

\section{Data management taxonomy}
\label{taxonomy}

We provide a taxonomy of tasks for multilayer network representation and manipulation, covering and extending operators found in major multilayer network software previously cited, 
which introduces a common basis for comparing various layer-supporting data structures.
Our taxonomy differs from the recently introduced visualization-centered one~\cite{VizBook}; our proposal focuses on data management operators, as typically provided by database management systems. 

A preliminary taxonomy can be seen in Table~\ref{tab:taxonomy}. Our operator examples consider a generic multilayer network model similar to the one by Kivelä et al.~\cite{Kivela2014}, for a multilayer network $M=(V_M,E_M,V,\textbf{L})$ with $d$ dimensions. We use $\delta$ to denote a dimension of the network, $v$ to denote a vertex, $l$ to denote a single layer and $\sigma$ to denote a predicate.

\begin{table}[ht!]
\centering
\caption{Preliminary taxonomy of layer-supporting data model operators}\label{tab:taxonomy}
\begin{tabular}{ll}
\hline\noalign{\smallskip}
\textbf{Operator class} & \textbf{Examples} \\
\noalign{\smallskip}\hline\noalign{\smallskip}
Layer definition 
    & create/delete-dimension$(\delta)$\\ 
    & create/delete-layer$(\delta)$\\
\noalign{\smallskip}\hline\noalign{\smallskip}
Layer manipulation
    & flatten-layer$(\delta,l_s,l_t)$\\ 
    & project-layer$(\delta,l_s,l_t)$\\
    & diff-layer$(\delta,l_s,l_t)$\\
\noalign{\smallskip}\hline\noalign{\smallskip}
Layer query
    & filter-layer$_{\sigma}(l)$\\
\noalign{\smallskip}\hline\noalign{\smallskip}
Data manipulation
    & add/update/remove-node$(v,l)$\\ 
    & add/update/remove-edge$(v_s,l_s,v_t,l_t)$\\
\noalign{\smallskip}\hline
\end{tabular}
\end{table}

Tasks in the layer definition category focus on redefining the layer structure by creating and deleting dimensions. Layer manipulation operators also create new layers (or views thereof), but they instead derive new layers based on existing layers' topological features, e.g. layer flattening, projection and difference. 

We consider layer query as a special class, as it arguably falls under both layer and data manipulation categories. Operators here essentially obtain a subset of the nodes and edges that satisfy a condition related to either attributes or topological features.

Finally, operators in the data manipulation category, similarly to manipulating a single-layer graph database, include adding and removing nodes or edges, while also considering the layers they associate to. 

\section{Benchmark}\label{bench}

Multilayer networks can be implemented using data structures as different as tensors, dictionary-based adjacency lists, tables in relational databases, or combinations of the previous, as with graph database systems. We perform a comparison of these structures on the operators in our taxonomy, aiming to discover processes in need of optimization. 
Our primary focus is on layer manipulation and query operators, as we expect their performance within different systems to vary with respect to the underlying data structure and whether that is layer-native. 

As an example, we consider the performance of the layer aggregation operator in \texttt{MuxViz}~\cite{MuxViz}, \texttt{multinet}~\cite{multinet} and \texttt{Pymnet}~\cite{Pymnet} libraries. 
This experiment is done for randomly generated two-layer Erd\H{o}s-R\'enyi multilayer networks, with network size (nodes associated to each layer) ranging between $100$--$100,000$ and average node degree $\approx 4$. 
Note that \texttt{multinet} and \texttt{MuxViz} are libraries for R, while \texttt{Pymnet} is a library for Python, which can slightly affect performance.

\begin{figure*}[ht!]
\centering
\begin{tikzpicture}
\begin{axis}[
    xlabel={Network size (\# nodes in layer)},
    ylabel={Performance (seconds)},
    xmin=0,xmax=45,
    ymode=log,
    xtick={0,5,10,15,20,25,30,35,40,45},
    xticklabels={100,200,500,1000,2000,5000,10000,20000,50000,100000},
    ymajorgrids=true,
    enlargelimits=0.05,
    legend pos=north west,
]
\addplot[color=blue,mark=square]
    coordinates {
    (0,0.009146690369)(5,0.01112508774)(10,0.0207760334)(15,0.03302717209)(20,0.06395173073)(25,0.1876349449)(30,0.3426980972)(35,0.7101509571)(40,2.096149683)(45,4.543964863)
    };
\addplot[color=red,mark=triangle]
    coordinates {
    (0,0.003238201)(5,0.006976843)(10,0.01918983)(15,0.04660797)(20,0.1053059)(25,0.3661661)(30,0.9207039)(35,2.077801)(40,6.166214)(45,12.92407)
    };
\addplot[color=black,mark=star]
    coordinates {
    (0,0.00267005)(5,0.0009961128)(10,0.0009889603)(15,0.001163006)(20,0.001376152)(25,0.002245903)(30,0.00427103)(35,0.006654024)
    };
\addplot[color=black,dotted]
    coordinates{
    (35,0.006654024)(40,0.006654024)(45,0.006654024)
    };
\node[] at (axis cs: 37.5,0.01) {memory overflow};
\legend{Pymnet,multinet,MuxViz}
\pgfplotsset{every tick label/.append style={font=\tiny}}
\end{axis}
\end{tikzpicture}
\caption{Comparison of three multilayer network libraries for performance of two layer network aggregation over network size.}
\label{fig:flatten_bench} 
\end{figure*}
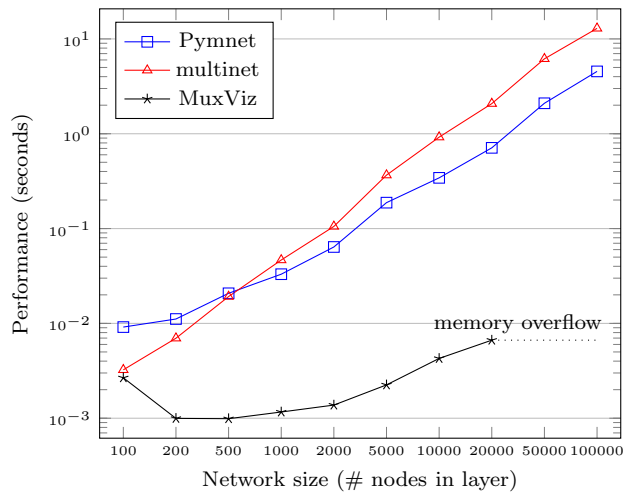

Fig.~\ref{fig:flatten_bench} confirms the aforementioned imbalance between data structure efficiency and scalability, as systems exhibit various behaviours related to their underlying data structures; notice that some systems can handle larger networks, but at the expense of efficiency. 

%
To expand our benchmark, we will also compare our findings with equivalent database models able to represent multilayer networks. Namely, we consider a relational database model representing each element of the multilayer network quadruple as its own relation and a graph database model operating on the network defined by $(V_M,E_M)$.

\section{Conclusion and future work}
Despite only dealing with preliminary results, we can already stress the importance of the chosen data structure, as this choice can determine which networks we can practically handle. These results should be complemented with a comparison of additional established multilayer network operators on both the aforementioned libraries and database models, in order to explore the strengths and weaknesses of various multilayer network management approaches.
Finally, future work includes defining a minimal set of necessary operators for  multilayer network data and a rule-based approach to transform multilayer networks into database models.


\begin{acknowledgements}
This work has been partly funded by eSSENCE, an e-Science collaboration funded as a strategic research area of Sweden, and the FRÖ program by the French Institute in Sweden.
\end{acknowledgements}

\bibliographystyle{spphys}       
\bibliography{bt-soft,bt-relgdb,bt-survey,bt-appl,bt-eer,bt-hin}  

\end{document}